\documentclass[12pt]{article}
\usepackage[pctex32]{graphics}
\begin{document}
\vskip 3.5 true cm
\begin{center}
{\LARGE  Coexistence of Anticipating Synchronization and Lag Synchronization in an Optical System}
\vskip 0.8 true cm
{{\sc Liang Wu} and {\sc Shiqun Zhu*}}
\vskip 0.2 true cm
{\it  China Center of Advanced Science and Technology (World Laboratory),\\ 
P. O. Box 8730, Beijing 100080, People's Republic of China}\\
{\it and}\\
{\it  Department of Physics, College of Sciences, Suzhou University,\\ 
Suzhou, Jiangsu 215006, People's Republic of China**}\\
\vspace{1.5cm}
\end{center}
\begin{abstract}
The chaotic synchronization between two bi-directionally coupled external cavity single-mode semiconductor lasers is investigated. Numerical simulation shows that anticipating synchronization and lag synchronization coexist in certain parameter regime. The anticipating time with different effects that were discussed quite differently in the previous theoretical analysis and experimental observation is determined by the involved parameters in the system.
\end{abstract}

\vspace{0.5cm}

PACS number(s): 05.45.Xt, 42.55.Px, 42.65.Sf

*Corresponding author

**Mailing address

E-mail: liangwu@suda.edu.cn or szhu@suda.edu.cn

\newpage
{\bf I. Introduction}
\quad

In recent years, much attention has been paid to the chaotic synchronization in nonlinear systems [1-3]. There is much prospect of chaotic synchronization being applied in various ways, especially in secure communication where many works have been conducted and a lot of progress has been made [2,4,7]. In last few years, the theory of chaotic synchronization are being utilized in the research of neural networks [5].

Chaotic synchronization was introduced as studying the dynamics of several classical models [1,2], such as Lorenz model. Then the experimental demonstration of the chaotic synchronization was given in the electrical systems and optical systems successively [2,4,7]. Particular emphasis is put upon the synchronization in the chaotic external cavity semiconductor lasers [6-13] because of their ability to generate high-dimensional chaos and their ease of operation. Several years ago, the investigation in chaotic synchronization was confined to $\textit{lag synchronizaiton}$ [6]. There is a retardation time ($\tau_c$) between two lasers' output due to the finite time for the light to travel from the master to the slave. Recently, it is found that there is also \textit{anticipating synchronization} [8-13] other than lag synchronization, the intensity of the slave is synchronized to the future intensity of the master. That means the slave can anticipate the dynamics of the master.

Although anticipating synchronization is counterintuitive, its existence can be illuminated both theoretically and experimentally [8-13]. Voss discovered anticipating synchronization in some simple models first [8]. Following this discovery, Masoller predicted anticipating synchronization in unidirectionally coupled lasers system [9] and showed that the anticipating time ($\tau_A$) should be equal to the difference between round-trip time of the light in the master's  external cavity and the retardation time, i.e., $\tau_a=\tau_c-\tau$. Recently Sivaprakasam reported the experimental demonstration of anticipating chaotic synchronization in a bi-directionally coupled external cavity semiconductor laser system [12] and found the anticipating time is irrespective of the external-cavity round trip time $\tau$, i.e. $\tau_a=\tau_c$. At present there is no fully theoretical explanation of the difference between the theoretical anticipating time and the experimental  observation [13].  

In this paper, the synchronization between two bi-directionally coupled chaotic external cavity semiconductor lasers is investigated. Numerical simulations show very interesting results. (1) Anticipating synchronization and lag synchronization coexist in a system. Sometimes the system is inclined to anticipating synchronization, sometimes to lag synchronization. And long-term behavior is decided by the involved parameters such as the coupling strengths and the feedback rate in the master.  
(2) As an important character of the system, the anticipating time is also determined by the parameters. The parametric space is divided into several zones. In one of these zones, anticipating time is the same as derived in theoretic analysis reported in previous papers. In another zone, anticipating time is in agreement with experimental observation.   

\vspace{0.5cm}
{\bf II. Theoretical Model}

The dynamics of two bi-directionally coupled single-mode semiconductor lasers with only the master laser subjected to external optical feedback can be described by the widely utilized Lang-Kobayashi equations [14]
\begin{eqnarray}
\frac{dE_m}{dt} & = & k_m(1+i\alpha_m)[G_m-1]E_m(t)+\eta_{sm} E_s(t-\tau_c)\times exp[-i(\omega_s \tau_c+\triangle\omega t)] \nonumber\\
& & +\gamma_m E_m(t-\tau)\times exp(-i\omega_m\tau)+\beta_m\xi_m(t)\\
\frac{dN_m}{dt} & = & \frac{j_m-N_m-G_m|E_m|^2}{\tau_{nm}}\\
\frac{dE_s}{dt} & = & k_s(1+i\alpha_s)[G_s-1]E_s(t) \nonumber+\eta_{ms} E_m(t-\tau_c) \nonumber\\
& & \times exp[-i(\omega_m \tau_c+\triangle\omega t)]+\beta_s\xi_s(t)\\
\frac{dN_s}{dt} & = & \frac{j_s-N_s-G_s|E_s|^2}{\tau_{ns}}
\end{eqnarray}

Where subscript m and s denote the master and the slave respectively. $E$ is the slowing varying complex field, and $N$ is the normalized carrier density. The second term on the right side in both (1) and (3) corresponds to the bi-directional coupling. $k$ is the cavity loss, $\alpha$ the linewidth enhancement factor. $G_i=N_i/(1+\epsilon_i|E_i|^2)$ is the optical gain where i stands for m or s and $\epsilon_i$ is the gain saturation coefficient. $\gamma$ is the feedback rate, $\eta$ is the coupling strength and subscript $sm$ indicates that the coupling is from the slave to the master, $ms$ indicates that the coupling is from the master to the slave, $\omega$ is the optical frequency without feedback,  $\triangle\omega=\omega_s-\omega_m$ is the frequency detuning between the lasers, $\tau$ is the external cavity round trip time in the master, and  $\tau_c$ is the time for the light to fly from the master to the slave. $\xi$ is independent complex Gaussian white noise, and $\beta_i$ measures the noise intensity. $j$ is the normalized injection current, and $\tau$ is the carrier lifetime. For simplicity, here we take $\beta_m=\beta_s= 0$  .

\vspace{0.5cm}
{\bf III. Coexistence of lag synchronization and anticipating synchronization}

According to the equations(1)-(4), we make the numerical simulation of the dynamics of two bi-directionally coupled semiconductor lasers. We figure out  $Q=\sigma^2(arctan(\frac{I_s}{I_m}))$ to represent the synchronization quality, where $I_m$ and $I_s$ are the field intensities of the master and slave respectively ($I_m = |E_m|^2,I_m = |E_m|^2$),  $\sigma^2$ is the variance calculation. Perfect synchronization is represented by $Q=0$, on the other hand a high variance $Q$ represents a poor synchronization. Because of the effect of anticipating synchronization or lag synchronization, we would derive a good synchronization represented by a low variance $Q$ if the master laser output is shifted relatively to the slave output by an appropriate time $\tau_s$. In this paper, $\tau_s$ takes $\tau_s>0$ if the master laser output is shifted forward relative to the slave output. In this case, a low variance $Q$ represents an anticipating synchronization. $\tau_s$ takes $\tau_s<0$ if the master laser output is shifted backward relative to the slave output, a low variance $Q$ in this case represents a lag synchronization.

We find that anticipating synchronization and lag synchronization coexist in the bi-directionally coupled semiconductor lasers system. The coexistence can be seen in fig.1. Fig.1 shows the dependence of the variance $Q$ on the master laser time shift $\tau_s$.
there is a low variance on each side. The low variance on the right side($\tau_s>0$) indicates the anticipating synchronization and the low variance on the left side($\tau_s<0$) indicates the lag synchronization. 
the coexistence is further illustrated in detail in fig.2(a-1),2(a-2). Fig.2(a-1) and 2(a-2) plot the time traces of the field intensities of the master(upper) and the slave(lower). Each trace can be naturally divided into many small zones, labelled by $m_1,m_2,m_3,m_4...$  in the master laser, $s_1,s_2,s_3,s_4...$in the slave laser. we detect the similarities between $s_1$ and $m_1$, $m_2$ and $s_1$, $s_2$ and $m_2$, $m_3$ and $s_2$, $s_3$ and $m_3$, $m_4$ and $s_3$.... The dynamics of the system of two mutual coupled semiconductor lasers is of lag synchronization as far as the similarities between $s_1$ and $m_1$, $s_2$ and $m_2$,   $s_3$ and $m_3$    are concerned, on the other hand, the dynamics is of anticipating synchronization as far as the similarities between $m_2$ and $s_1$, $m_3$ and $s_2$, $m_4$ and $s_3$ are concerned. Thus, the two kinds of synchronization coexist with each other.
In addition, the similarities between $m_i$ and $s_i$, $s_i$ and $m_{i+1}$ fluctuate. The fluctuations indicate that sometimes the system behavior is inclined to anticipating synchronization orientation, sometimes is inclined to lag synchronization orientation. Six traces in fig.2 share one dynamical trajectory. Fig.2(a-1) and 2(a-2) plot the intensities in the master and slave respectively in the range 250-300ns; fig.2(b-1) and 2(b-2) plot in the range 410-450ns, typically showing lag synchronization orientation; fig.2(c-1) and 2(c-2) plot in the range 900-930ns, typically showing anticipating synchronization orientation. 

Whether the long-term system behavior is anticipating or lag is decided by the involved parameters. In this paper the coupling strengths in both directions and  the feedback rate in the master are discussed. We use  $\Delta Q$ to indicate the orientation of the synchronization in a sufficient long term.
\begin{eqnarray}
Q_a & = & Q_{min} \quad (\tau_s>0)\\
Q_l & = & Q_{min} \quad (\tau_s<0)\\
\Delta Q & = & Q_a - Q_l 
\end{eqnarray}
where a and l denote anticipating synchronization and lag synchronization respectively. 
The anticipating synchronization orientation is charactered by positive $\Delta Q$ ($\Delta Q>0$), while lag synchronization orientation is charactered by negative $\Delta Q$ ($\Delta Q<0$).

In fig.3(a) we plot $\Delta Q$ as a function of the feedback rate in the master ($\gamma_m$) and the coupling strength from the slave to the master ($\eta_{sm}$). The coupling strength from the master to the slave $\eta_{ms}=5.0$. We can find  $\Delta Q>0$ in two regions labelled by $I_1$, $I_2$, where the long-term system behavior is inclined to anticipating synchronization orientation. In another region labelled by II, $\Delta Q<0$ indicates the long-term system behavior is inclined to lag synchronization orientation. In region III, the long-term system dynamics is adiaphorous, represented by $\Delta Q\approx0$.

In fig.3(b) we plot $\Delta Q$ as a function of the couplings in both directions($\eta_{ms}$,$\eta_{sm}$). We find that almost all  the plane has $\Delta Q\leq0$, which indicates the long-term system behavior is inclined to lag synchronization in most cases. We mention that  only the master has feedback which break the system's symmetry. The symmetry is considered in the following section.

\vspace{0.5cm}
{\bf IV. Anticipating time}

There is a difference between the theoretical anticipating time reported in previous papers and the experimental observation. Our numerical simulation shows that the two cases both occur in bi-directionally coupled chaotic external cavity semiconductor lasers system. In our numerical simulation, we take $\tau_c=3.5ns, \tau=6.7ns$. Fig.3(c) plots the calculated anticipating time ($\tau_a=\tau_s,\tau_s>0$) as the function of the feedback rate in the master ($\gamma_m$) and the coupling strength from the slave to the master ($\eta_{sm}$).
In terms of $\tau_a$ one can clearly distinguish two regions. In the region labelled by I,  $\tau_a$ is about 3.5ns, in agreement with experimental anticipating time ($\tau_a=\tau_c=3.5ns$). In the region labelled by II in fig.3(c), $\tau_a$ is about 3.2ns, corresponding to previous theoretical reports ($\tau_a=\tau-\tau_c=3.2ns$).  

Fig.3(d) shows the dependence of anticipating time on the couplings in both directions. The two regions, one for $\tau_a=\tau_c=3.5ns$, the other for $\tau_a=\tau-\tau_c=3.2ns$ could also be distinguished clearly.

With respect to the anticipating time, an important characteristic scalar, the system is symmetric.($\eta_{ms}$,$\eta_{sm}$)=(a,b) and ($\eta_{ms}$,$\eta_{sm}$)=(b,a) give the same $\tau_a$. The symmetry tell that the two lasers should be regarded as segments of a whole system. Two lasers participate in the dynamical evolution jointly, which could  also be detected in fig.2. We mention that the bi-directional coupling leads to the symmetry. And the symmetry indicates that the couplings in two directions are equivalent.

\vspace{0.5cm}
{\bf Acknowledgements}

It is a pleasure to thank Weijian Gao for his many helpful discussions of numerical calculations. The financial supports from the National Natural Science Foundation of China (Grant No. 19874046) and the Natural Science Foundation of Jiangsu Province (Grant No. BK2001138) are gratefully acknowledged.

\newpage
\begin{center}{Figure Captions}\end{center}
{\bf Fig. 1.} Variance $\sigma^2(arctan(\frac{I_s}{I_m}))$ as a function of the shift time $\tau_s$. The parameters are $k_m=k_s=500$ns, $\alpha_m=\alpha_s=3$, $\epsilon_m=\epsilon_s=0.1$,$\tau_{nm}=\tau_{ns}=1.0$ns, $j_m=j_s=1.01$, $\omega_m=\omega_s=0.3$rad$\cdot$ ns$^{-1}$, $\gamma_m=2$ns$^{-1}$, $\eta_{ms}=5$ns$^{-1}$, $\eta_{sm}=3$ns$^{-1}$, $\tau=6.7$ns, $\tau_c=3.5$ns
\\
\\
{\bf Fig. 2.} The time traces of the master laser field intensity (a-1)(b-1)(c-1) and the slave laser field intensity (a-2)(b-2)(c-2). (a-1)(a-2) range 250-300ns; (b-1)(b-2) range 410-450ns; (c-1)(c-2) range 900-930ns. $\gamma_m=10$ns$^{-1}$, $\eta_{ms}=10$ns$^{-1}$, $\eta_{sm}=2$ns$^{-1}$ [all other parameters are the same as fig.1].
\\
\\
{\bf Fig. 3.} (a) The calculated $\Delta Q$ as a function of the feedback rate in the master $\gamma_m$ and the coupling strength from the master to the slave $\eta_{sm}$. $\eta_{sm}=5$ns$^{-1}$. (b) The calculated $\Delta Q$ as a function of the couplings in both directions ($\eta_{ms}$, $\eta_{sm}$). $\gamma_m=5$ns$^{-1}$. (c) The calculated anticipating time as a function of $\gamma_m$ and $\eta_{sm}$. $\eta_{sm}=5$ns$^{-1}$. (d) The calculated anticipating time as a function of $\eta_{ms}$ and $\eta_{sm}$. [all other parameters are the same as fig.1].

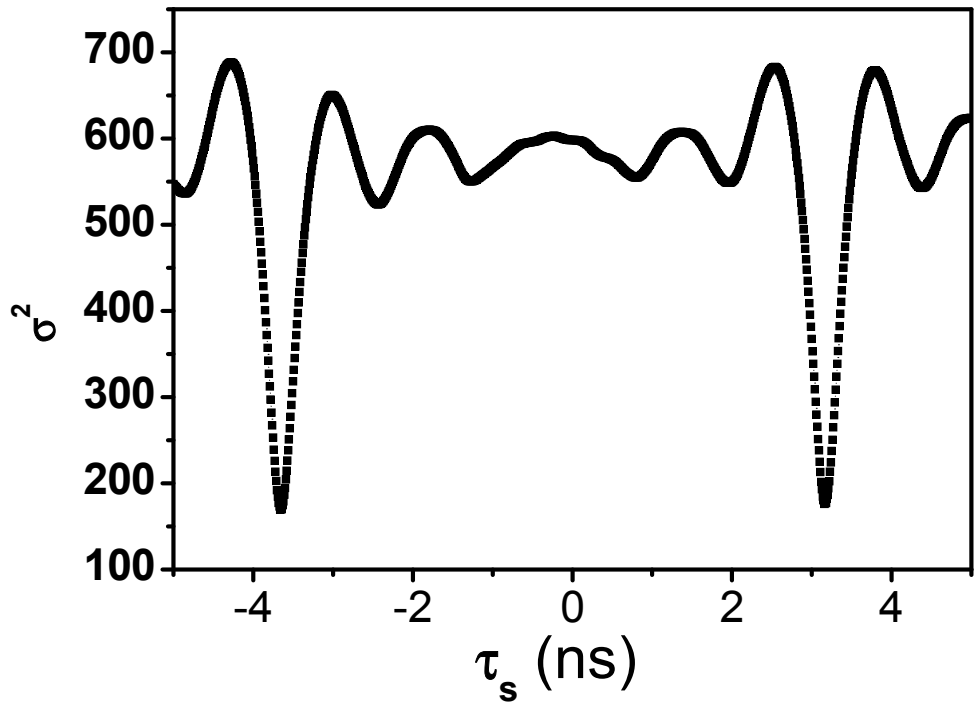
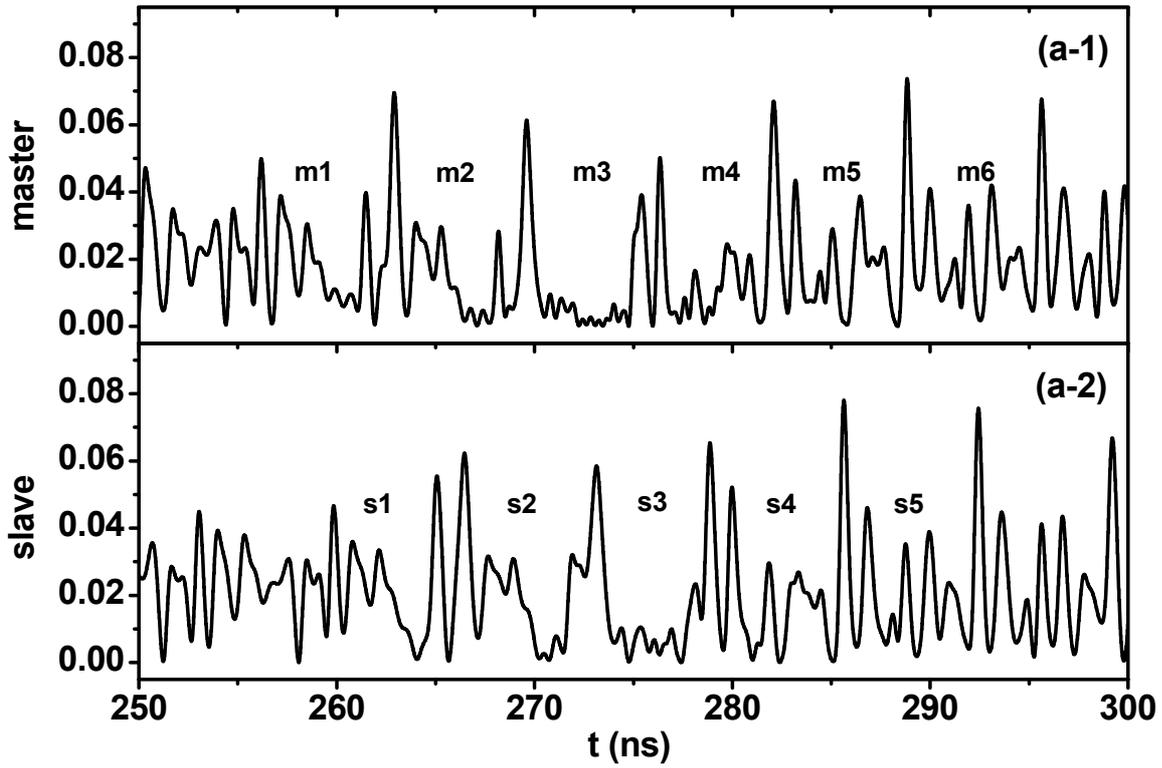
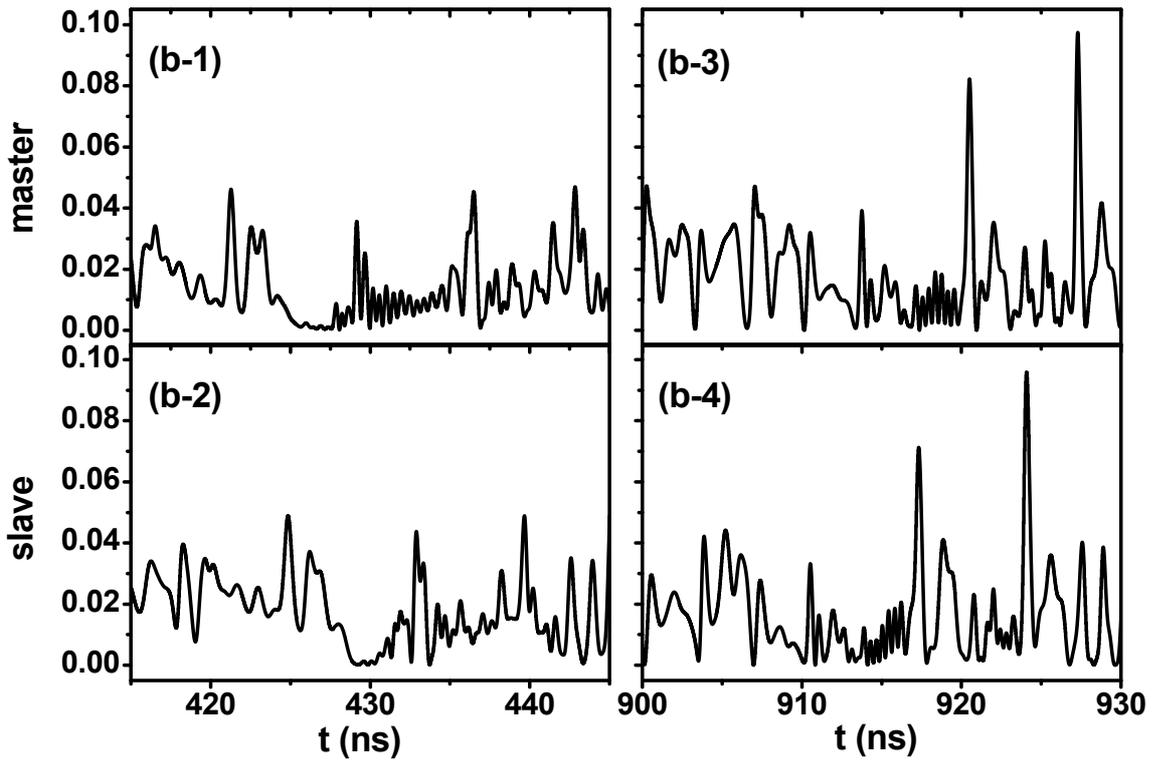
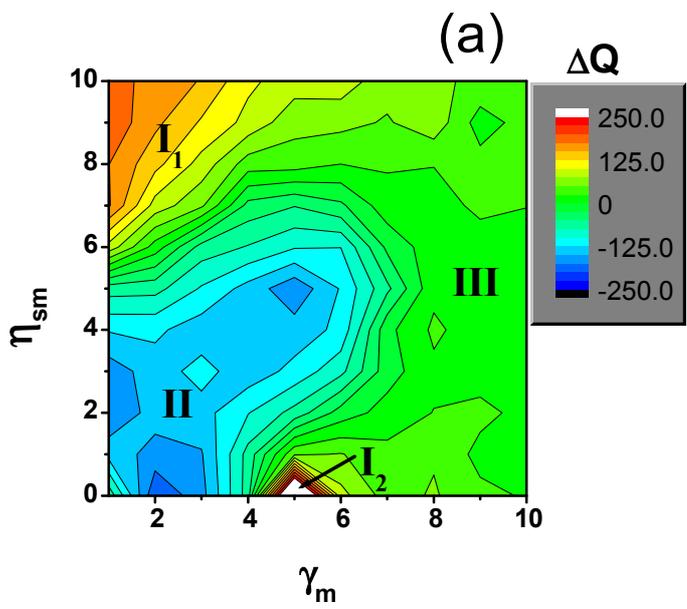
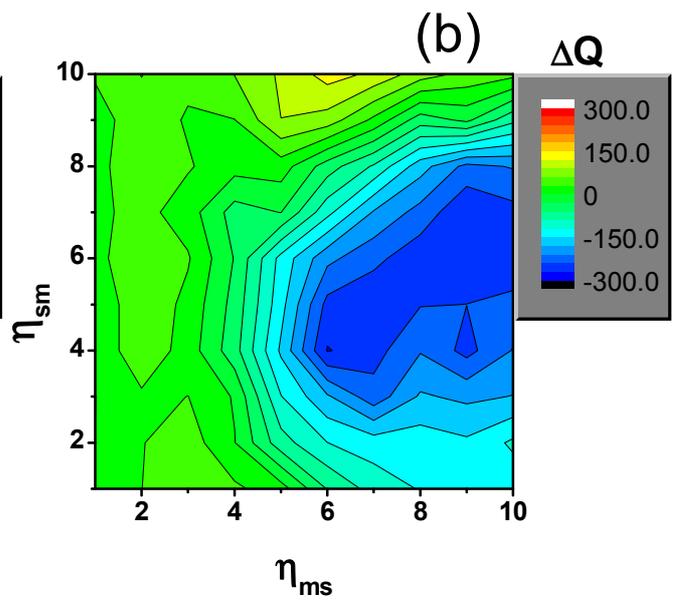
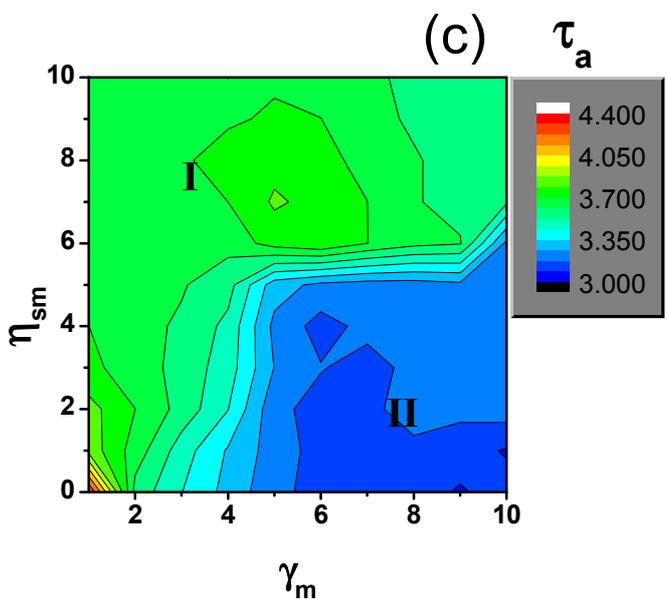
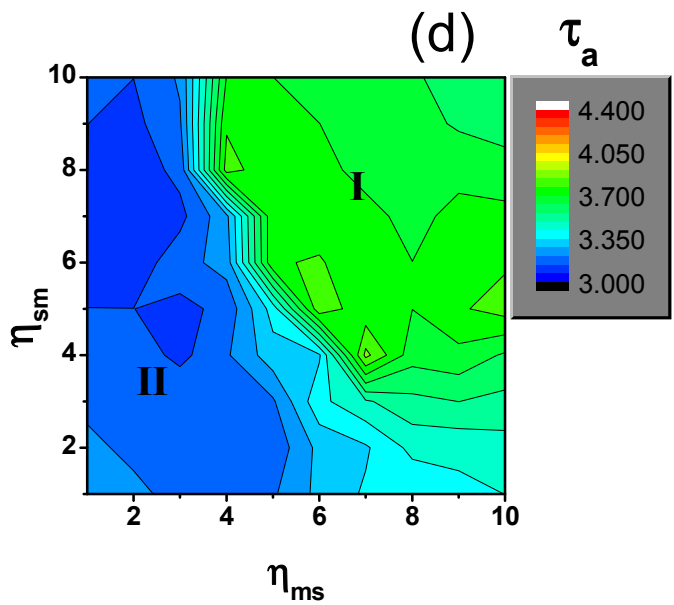
\end{document}